\documentclass[manuscript]{aastex}

\shorttitle{Lucifer Near-IR Tomography of NGC 1569}
\shortauthors{Pasquali et al.}

\begin{document}

\title{Infrared Narrow-Band Tomography of the Local Starburst \\ NGC 1569
with LBT/LUCIFER}

\author{A. Pasquali\altaffilmark{1}, A. Bik, S. Zibetti\altaffilmark{2}}
\affil{Max-Planck Institut f\"ur Astronomie, K\"onigstuhl 17,
    69117 Heidelberg, Germany}

\author{N. Ageorges}
\affil{Max-Plack-Institut f\"ur extraterrestrische Physik,
Giessenbachstrasse, 85748 Garching, Germany}

\author{W. Seifert}
\affil{Landessternwarte Heidelberg, K\"onigstuhl, 69117 Heidelberg, Germany}

\author{W. Brandner, H.-W. Rix}
\affil{Max-Planck-Institut f\"ur Astronomie, K\"onigstuhl 17,
    69117 Heidelberg, Germany}

\author{M. J\"utte, V. Knierim}
\affil{Ruhr-Universit\"at, Universit\"atstrasse 150, 44780 Bochum, Germany}

\author{P. Buschkamp}
\affil{Max-Plack-Institut f\"ur extraterrestrische Physik,
Giessenbachstrasse, 85748 Garching, Germany}

\author{C. Feiz}
\affil{Landessternwarte Heidelberg, K\"onigstuhl, 69117 Heidelberg, Germany}

\author{H. Gemperlein}
\affil{Max-Plack-Institut f\"ur extraterrestrische Physik,
Giessenbachstrasse, 85748 Garching, Germany}

\author{}
\affil{}

\author{A. Germeroth}
\affil{Landessternwarte Heidelberg, K\"onigstuhl, 69117 Heidelberg, Germany}

\author{R. Hofmann}
\affil{Max-Plack-Institut f\"ur extraterrestrische Physik,
Giessenbachstrasse, 85748 Garching, Germany}

\author{W. Laun}
\affil{Max-Planck Institut f\"ur Astronomie, K\"onigstuhl 17,
    69117 Heidelberg, Germany}

\author{R. Lederer}
\affil{Max-Plack-Institut f\"ur extraterrestrische Physik,
Giessenbachstrasse, 85748 Garching, Germany}

\author{M. Lehmitz, R. Lenzen, U. Mall}
\affil{Max-Planck-Institut f\"ur Astronomie, K\"onigstuhl 17,
    69117 Heidelberg, Germany}

\author{H. Mandel, P. M\"uller}
\affil{Landessternwarte Heidelberg, K\"onigstuhl, 69117 Heidelberg, Germany}

\author{V. Naranjo}
\affil{Max-Planck-Institut f\"ur Astronomie, K\"onigstuhl 17,
    69117 Heidelberg, Germany}

\author{K. Polsterer}
\affil{Ruhr-Universit\"at, Universit\"atstrasse 150, 44780 Bochum, Germany}

\author{A. Quirrenbach, L. Sch\"affner}
\affil{Landessternwarte Heidelberg, K\"onigstuhl, 69117 Heidelberg, Germany}

\author{C. Storz}
\affil{Max-Planck-Institut f\"ur Astronomie, K\"onigstuhl 17,
    69117 Heidelberg, Germany}

\author{P. Weiser}
\affil{Fachhochschule f\"ur Technik und Gestaltung, Speyerer Strasse 4, 68163 Mannheim,
Germany}

\altaffiltext{1}{Astronomisches Rechen Institut, M\"onchhofstrasse 12 - 14,
69120 Heidelberg, Germany}

\altaffiltext{2}{Dark Cosmology Centre, Niels Bohr Institute - University of Copenhagen,
Juliane Maries Vej 30, DK-2100 Copenhagen, Denmark} 

\begin{abstract}
We used the near-IR imager/spectrograph LUCIFER mounted on the Large Binocular
Telescope (LBT) to image, with sub-arcsec seeing, the local dwarf 
starburst NGC~1569 in the JHK bands
and HeI 1.08$\mu$m, [FeII] 1.64$\mu$m and Br$\gamma$ narrow-band filters.
We obtained high-quality spatial maps of HeI 1.08$\mu$m, [FeII] 1.64$\mu$m and 
Br$\gamma$ emission across the galaxy, and used them together with $HST$/ACS images 
of NGC~1569 in the H$\alpha$ filter to derive the two-dimensional spatial map 
of the dust extinction and surface star formation rate density. 
We show that dust extinction (as derived from the H$\alpha$/Br$\gamma$ flux ratio) is 
rather patchy and, on average, higher in the North-West (NW) portion of the galaxy
[E$_{\rm g}$(B-V) $\simeq$ 0.71 mag] than in the South-East [E$_{\rm g}$(B-V) $\simeq$ 
0.57 mag]. Similarly, the surface density of star formation rate 
(computed from either the dereddened H$\alpha$ or dereddened Br$\gamma$ image)  
peaks in the NW region of NGC 1569, reaching a value of about
4 $\times$ 10$^{-6}$ M$_{\odot}$~yr$^{-1}$~pc$^{-2}$. The total star formation rate 
as estimated from the integrated, dereddened H$\alpha$ (or, alternatively, Br$\gamma$) 
luminosity is about 0.4 M$_{\odot}$~yr$^{-1}$, and the total supernova rate from the integrated, 
dereddened [FeII] 1.64$\mu$m luminosity is about 0.005 yr$^{-1}$ (assuming a distance of 3.36 Mpc).
The azimuthally averaged [FeII] 1.64$\mu$m/Br$\gamma$ flux ratio is larger at the edges of
the central, gas-deficient cavities (encompassing the super star clusters A and B) and in
the galaxy outskirts. If we interpret this line ratio as the ratio between the average past
star formation (as traced by supernovae) and on-going activity (represented by OB stars able 
to ionize the interstellar medium), it would then indicate that star formation has been 
quenched within the central cavities and lately triggered in a ring around 
them. The number of ionizing hydrogen and helium photons as computed from the integrated,
dereddened H$\alpha$ and HeI 1.08$\mu$m luminosities suggests that the latest burst
of star formation occurred about 4 Myr ago and produced new stars with a total mass 
of $\simeq$1.8 $\times$ 10$^6$ M$_{\odot}$.

\end{abstract}


\section{Introduction}
Understanding the star formation activity and history of local dwarf
galaxies is of high astrophysical relevance. In fact, they represent ideal 
laboratories to study how star formation occurs in low-metallicity environments
that, at the same time, do not show any global, ordered kinematic structure as in disk galaxies. As such,
dwarf galaxies allow us to study how stellar feedback alone affects the physics, 
kinematics and chemical enrichment of the interstellar medium. Nearby, star-forming 
dwarf galaxies that are metal-poor and gas-rich are often compared
to the first galaxies in the early Universe (see Izotov \& Thuan 1999), which were 
the first structures to collapse from primordial density fluctuations, and gave rise 
to larger systems through mergers (White \& Frenk 1991; Kauffmann, White \& Guiderdoni 1993; 
Cole et al. 1994). They are also referred to as the possible local counterparts of starbursting
galaxies that, at moderate redshift, dominate the faint galaxy counts at 
blue wavelengths (Broadhurst, Ellis \& Glazebrook 1992; Lilly et al. 1995).
\par
The nearby galaxy NGC~1569 is considered to be the archetype starbust
dwarf galaxy for several reasons. First of all, it is gas-rich
and relatively metal-poor: at the revised distance of 3.36 Mpc
(Grocholski et al. 2008, so that M$_B$ = -18), its HI and dynamical masses
are $\sim$2 $\times$ 10$^8$ M$_{\odot}$ and 5.1 $\times$ 10$^8$ 
M$_{\odot}$, respectively (adapted from Reakes 1980). Its gas-phase and stellar 
metallicities are estimated to be 12 $+$ log(O/H) $\simeq$ 8.3 (Calzetti, Kinney 
\& Storchi-Bergmann 1994; Gonz\'alez Delgado et al. 1997, Kobulnicky \&
Skillman 1997) and $Z \sim$ 0.1 - 0.2 $Z_{\odot}$ (Grocholski et al. 2008; Aloisi et 
al. 2001), respectively. Star formation has taken place in NGC~1569 for 
a Hubble time, although with different intensities as obtained by Vallenari \& 
Bomans (1996), Greggio et al. (1998), Aloisi et al. (2001) and Angeretti et al. (2005) 
using color-magnitude diagrams of resolved stars. 
The recent studies by Angeretti et al. (2005) and Grocholski et al. (2008) show 
that the star formation activity of NGC~1569 started about 10 Gyr ago and proceeded 
until $\sim$1 Gyr ago presumably at a constant, very low rate. In the last 1  Gyr, 
however, NGC~1569 experienced at least three major bursts of star formation: 
{\it i)} the older one started about 1 Gyr ago and ended $\sim$100 Myr ago with an average
star formation rate ($<$SFR$>$) of about 0.04 M$_{\odot}$~yr$^{-1}$; {\it ii)} 
the intermediate episode was triggered about $\sim$ 100 Myr ago and lasted up to 30 Myr 
ago with $<$SFR$>$ $\sim$ 0.08 M$_{\odot}$~yr$^{-1}$, and {\it iii)} the younger burst begun
$\sim$ 27 Myr ago and ended about 8 Myr ago with $<$SFR$>$ $\sim$ 0.3 M$_{\odot}$~yr$^{-1}$
(Angeretti et al. 2005).
According to Aloisi et al. (2001), the spatial distribution of stars across the galaxy changes 
with their age, so that stars younger than 50 Myr are more centrally concentrated, intermediate age 
stars (50 Myr - 1 Gyr) are uniformly distributed, and older stars are mostly found 
in the outskirts of NGC~1569. Angeretti et al. (2005) suggested that this episodic star formation
(in the last 1 Gyr) could be ascribed to the gravitational interactions of NGC~1569
with an HI cloud ($\sim$ 7 $\times$ 10$^6$ M$_{\odot}$ in mass) a few kpc away yet
connected to the galaxy by an HI bridge (cf. Stil \& Israel 1998).
\par
The presence of several HII regions (Waller 1991) and the spatially extended H$\alpha$
emission indicate that NGC~1569 has formed new stars rather recently.
Hunter \& Elmegreen (2004) used the spatially integrated H$\alpha$ luminosity
(corrected for reddening) to derive a SFR of $\sim$ 0.6 M$_{\odot}$~yr$^{-1}$
(adjusted to a distance of 3.36 Mpc). The morphology of the extended H$\alpha$
is rather complex and characterized by arcs, filaments and four
large-scale superbubbles (Hunter et al. 1993; Heckman et al. 1995; Martin 1998;
Westmoquette et al. 2008). 
The latter are seen to expand at typical velocities of $\sim$100 km~s$^{-1}$ 
which imply dynamical ages $<$ 25 Myr, consistent with the most recent burst of star 
formation derived by Angeretti et al. (2005; Heckman et al. 1995; Westmoquette, Smith \& Gallagher 2008).
The expansion of these superbubbles is most likely powering a galactic outflow throughout
the disk whose direction is approximately perpendicular to the
inclined and flattened HI disc of NGC~1569. The presence of an outflow of ionized gas 
is also supported by the high [SII]/H$\alpha$ ratios measured within the superbubbles 
and indicative of shocks (Westmoquette, Smith \& Gallagher 2008). The arcs of ionized
gas detected in H$\alpha$ and X-ray in the superbubbles suggest that the hot
gas is still confined within the superbubbles, although it is moving at the escape
velocity of NGC~1569. Thus, at present day, the galactic outflow of NGC~1569 does not 
appear to be in a steady state and to be able to induce mass loss from the galaxy
(Westmoquette, Smith \& Gallagher 2008).
\par
NGC~1569 is special also for another reason: it hosts  two of the nearest
super-star clusters known (SSCs A and B, Arp \& Sandage 1985). SSCs are massive star 
clusters few Myr to several hundred Myr old, believed to be the ancestors 
of globular clusters. In fact, they are as extended and massive as globular clusters 
and, after $\sim$10 Gyr of passive evolution, they can match the luminosity range spanned 
by globular clusters today (cf. van den Bergh 1995; Meurer 1995). 
They are typically observed in interacting/merger systems 
(see Whitmore et al. 1993) and starburst galaxies (see Meurer et al. 1995). In NGC~1569,
SSC A has been resolved into two components (A1 and A2, De Marchi et al. 1997), of which
A2 may be the host of young O and Wolf-Rayet stars and A1 together with SSC B may be 
dominated by older red supergiants (Origlia et al. 2001). Gonz\'alez Delgado et al. (1997) 
suggested that SSCs A and B have possibly undergone sequential star formation, with 
two major bursts of star formation about 3 and 9 Myr ago. The authors also found a deficit of 
ionized gas around SSCs A and B which may have been created by the stellar winds and supernova 
explosions of the older burst removing the local gas. According to the measurements of Ho \& Filippenko
(1996) SSC A is as massive as $\sim$4 $\times$ 10$^5$ M$_{\odot}$ (adjusted to a distance
of 3.36 Mpc). 
\par
Age dating of star clusters and resolved stars strongly depends on dust extinction.
Studies of the stellar content of NGC~1569 based on long-slit spectroscopy show that 
the intrisic dust extinction varies across NGC~1569 by few tenths of a dex, and peaks at 
SSCs A and B (Gonz\'alez Delgado et al. 1997; Origlia et al. 2001) and in the North-West portion of 
the galaxy (Kobulnicky \& Skillman 1997). The spatial coverage of long-slit spectroscopy 
is coarse, however, and no uniform and contiguous two-dimensional map of the dust extinction 
in NGC~1569 is available in the literature. Clearly, such a 2D map helps in reaching a
higher accuracy in the determination of the star formation history and its associated 
star formation rate across NGC~1569. For this reason, we have taken advantage of the large
field of view of LUCIFER (LBT NIR spectroscopic Utility with Camera and Integral-Field Unit
for Extragalactic Research) mounted on the Large Binocular Telescope (LBT) to perform
deep imaging in the Br$\gamma$ filter and in the K band. Our aim is to construct 
the 2D spatial map of dust extinction across NGC~1569 from the Br$\gamma$ and H$\alpha$ 
(obtained with the Advanced Camera for Surveys on HST) images of the galaxy and to derive the 
2D spatial distribution of star formation rate density. We also observed the HeI 1.08$\mu$ and 
[FeII] 1.64$\mu$ lines and their broad-band continua in order to estimate the number of 
massive young stars and the supernova rate in NGC~1569, respectively. 
The observations and data reduction are described in Sect. 2. The spatial distribution of the
color excess E$_{\rm g}$(B-V) associated with the gas and the spatial distribution of the star 
formation rate density across NGC~1569 are derived in Sect. 3 and 4, respectively, 
while the properties of the [FeII] 1.64$\mu$m emission are analyzed in Sect. 5. 
Conclusions follow in Sect. 6.

\section{Observations}
The observations presented here were carried out with the LBT, located on Mount Graham,
Arizona (Hill, Green \& Slagle 2006). NGC~1569 was observed during the science commissioning 
of LUCIFER (Ageorges et al. 2010; Seifert et al. 2010), with the N3.75 camera which provides 
a 4$' \times$ 4$'$ 
field of view at an angular resolution of 0.12$''$/pixel (equivalent to 1.95 pc per pixel 
at the adopted distance of 3.36 Mpc). We imaged NGC~1569 between
September and November 2009 in the JHK broad bands and in the narrow-band HeI 1.08$\mu$m,
[FeII] 1.64$\mu$m and Br$\gamma$ filters, with a seeing $\leq$ 0.5$''$ (i.e. 8 pc) throughout
the observing runs. 
Although the exposure times in the broad-band and Br$\gamma$ filters were estimated from
the available 2MASS photometry and $HST$/ACS F658N (H$\alpha$) image via the LUCIFER Exposure
Time Calculator, the effective exposure times were largely dictated by the available 
amount of time during the LUCIFER commissioning. The exposure time for each of the HeI 1.08$\mu$m
and [FeII] 1.64$\mu$m filters was initially set to 1 hour (following Labrie \& Pritchet 2006),
but it was later cut down to 30 min for the HeI 1.08$\mu$m.
The final exposure times on source achieved for the near-IR dataset are 
listed in Table 1, together with the imaging taken in the F606W (broad V) and 
F658N (H$\alpha$) filters with $HST$/ACS as part of program ID 10885 (PI Aloisi).
\par
The near-IR images were reduced with standard IRAF\footnote{IRAF is distributed by the
National Optical Astronomy Observatory, which is operated by the Association of
Universities for Research in Astronomy, Inc., under cooperative agreement with the
National Science Foundation.} routines. They were corrected for bias and flat-fielding,
and their astrometric solution was derived from the field stars in common with 2MASS.
After background subtraction, the images taken in the same filter were corrected for
geometric distortion and then combined together
on the basis of their World Coordinate System with a weighted mean. The final mosaiced frames 
in the broad band filters were flux-calibrated, at a 10\% accuracy, using the 2MASS magnitudes 
of the stars in the field of NGC~1569.
\par\noindent
After convolution to the same PSF (FWHM = 0.5$''$) the continuum emission was subtracted from 
each narrow-band image by scaling the closest broad-band image (J for HeI 1.08$\mu$m, H for [FeII] 
1.64$\mu$m and K for Br$\gamma$) so that field stars would have had the same integrated count 
rate in both the narrow and broad band filters. This scaling was applied to the broad band 
zero points in order to calibrate the pure emission line images in flux, with an achieved 
accuracy of 20\% (estimated from the scatter in the zero point derived from different
stars).
\par
The optical HST/ACS images were retrieved from the HST archive already reduced. The
subtraction of the continuum emission (the F606W image) from the F658N frame and
the flux calibration of the pure H$\alpha$ emission were performed as for the near-IR data, 
reaching a flux accuracy of 20\%. Such a low accuracy in the flux calibration of the H$\alpha$ 
is mostly due to the broad F606W filter which makes it impossible to account for color terms.
The H$\alpha$-emission image was corrected for a 5\% flux contribution from the [NII] $\lambda$6548,6584
lines (Moustakas \& Kennicutt 2006), and then convolved with a Gaussian function to have the same
stellar PSF as the Br$\gamma$-emission image, resampled to achieve the same pixel scale 
as the LUCIFER N3.75 camera, and aligned to the Br$\gamma$-emission image.
\par
 Most of the analysis presented in the following sections requires
  or involves the production of spatially resolved maps of flux ratios
  in different passbands. In order to ensure a sufficient S/N for the
  largest number of pixels while retaining the best possible spatial
  resolution, we adopt the approach introduced by Zibetti, Charlot \&
  Rix (2009) using the adaptive smoothing code ADAPTSMOOTH (Zibetti
  2009). We first smooth each image individually by taking the median
  flux among the pixels inside a top-hat circular kernel, whose size
  is determined as the minimum that allows to obtain a S/N$\ge 10$
  (background dominated noise is assumed). The spatial match between
  different bands is obtained by requiring that the kernel size at
  each pixel location ensures a minimum S/N of 10 in all images (see
  Zibetti 2009 for details). The smoothing procedure also results in a
  first S/N cut: all pixels for which sufficient S/N of 10 cannot be
  reached in all images even when the maximum kernel size of 27 pixels
  is used are flagged. A further S/N cut is applied by excluding all
  pixels whose flux (in the smoothed image) is less than 3$\sigma$ of
  the large scale background fluctuations as measured from randomly
  placed boxes of a few hundred pixels on a side.

\section{The spatial map of dust extinction}
The line-emission images of NGC~1569 are shown in Fig. 1, already flux-calibrated but not
corrected for dust extinction. Fluxes, in units of erg~cm$^{-2}$~s$^{-1}$, are displayed 
in logarithmic scale across an area
of 1.5 kpc $\times$ 1 kpc at the adopted distance of 3.36 Mpc. We clearly detect the base 
of the north-western arm as well as ``cavities'' at the center of the galaxy where SSCs A and B
reside, and corresponding to the cluster surroundings where Gonz\'alez Delgado et al.
(1997) found a deficit of ionized gas. These regions were flagged as underflow pixels by our
smoothing procedure.
The S/N cut used in our image smoothing results in a limiting flux of $\approx 10^{-17}$,
$\approx 10^{-17.5}$, $\approx 10^{-18}$ and $\approx 10^{-18.5}$ erg~cm$^{-2}$~s$^{-1}$
for H$\alpha$, HeI 1.08$\mu$m, [FeII] 1.64$\mu$m and Br$\gamma$, respectively
(corresponding to the lower limits of the flux ranges displayed in Fig. 1).
The morphologies of the H$\alpha$ and HeI 1.08$\mu$m 
emissions, which trace the location of young and massive stars, are rather similar.
They exhibit two major peaks along the major axis of NGC~1569, one on each side of the 
cavities and with the brighter peak located in the North-West (NW) half of the galaxy.
Relatively high fluxes are also detected along nearly the full perimeter of the cavities. 
The Br$\gamma$ emission is morphologically consistent with that of the H$\alpha$ although 
it seems to be relatively weaker along the eastern rim of the cavities. Finally, 
the [FeII] 1.64$\mu$ emission, which is typically due to shocks induced by supernova 
explosions (but see Sect. 5), is as diffuse as the H and He emissions. The [FeII] flux peaks 
only in the NW half of NGC~1569 and stays relatively high along the southern rim of the cavities.
The very bright compact source in the eastern part of NGC~1569 is a supernova remnant
(cf. Labrie \& Pritchet 2006). 
\par
We integrate the observed H$\alpha$ and Br$\gamma$ maps of NGC~1569 over an ellipse of
semi-major axis $a$ = 37$''$ (606 pc) and semi-minor axis $b$ = 17$''$ (274 pc) which does
not include the extended H$\alpha$ superbubbles, where shocks due to the galactic outflow
seem to be important. The depth of our near-IR imaging is not sufficient to allow us
to detect these extended superbubbles. Pixels below our S/N cut are excluded from the
integration.\footnote{We compared the resulting total fluxes with those obtained
from the same area prior applying ADAPTSMOOTH to find negligible differences, at the 1\% - 2\%
level.} We use the resulting total H$\alpha$ and Br$\gamma$ fluxes to derive the color excess
(intrinsic $+$ foreground) averaged across the galaxy as in Calzetti et al. (1996):

\begin{equation}
{\rm <E_{\rm g}(B-V)>  =  \frac{-log(R_{\rm obs}/R_{int})}{0.4[\kappa(\lambda_{\rm H\alpha}) - 
\kappa(\lambda_{\rm Br\gamma})]}}
\end{equation}

\par\noindent
where R$_{\rm obs}$ is the observed H$\alpha$/Br$\gamma$ flux ratio, R$_{int}$ is the intrisic
H$\alpha$/Br$\gamma$ flux ratio (102.8) computed for Case B (i.e. the gaseous clouds are optically 
thick to line photons) assuming a temperature of 10,000 K and a density of 100 cm$^{-3}$ (Osterbrock 1989).
The extinction curve $\kappa(\lambda)$ is adopted from Fitzpatrick (1999), with 
$\kappa(\lambda_{\rm H\alpha})$ = 2.535 and $\kappa(\lambda_{\rm Br\gamma})$ = 0.363.
We thus obtain ${\rm <E_{\rm g}(B-V)>}$ = 0.58 mag, consistent with Calzetti et al. (1996), and an
intrisic  ${\rm <E_{\rm g}^{i}(B-V)>}$ = 0.07 mag assuming a foreground Galactic color excess 
${\rm E_{\rm G}(B-V)}$ = 0.51 mag (Burstein \& Heiles 1982; Calzetti et al. 1996). 
Since the intrinsic color excess on the stellar continuum is
${\rm <E_{\rm S}^{i}(B-V)>}$ = 0.44${\rm <E_{\rm g}^{i}(B-V)>}$, we estimate the total (intrinsic $+$
foreground) color excess on the stellar continuum to be ${\rm <E_{\rm S}(B-V)>}$ = 0.54 mag, 
consistent with the values derived by Angeretti et al. (2005) and Grocholski et al. (2008) from the 
color-magnitude diagram of resolved stars.

\par 
We apply Eq. (1) to each pixel in the observed H$\alpha$ and Br$\gamma$ maps in order to
derive the spatial distribution of E$_{\rm g}$(B-V) (intrinsic $+$ foreground) across NGC~1569.
This is shown in Fig. 2; we clearly see that the color excess is higher ($>$ 0.6 mag) where the H 
and He emission fluxes are higher, and the NW half of the galaxy is more extinguished than the South-East 
[where E$_{\rm g}$(B-V) $<$ 0.6 mag]. The distribution of the pixel color
excess as derived from Fig. 2 peaks at about E$_{\rm g}$(B-V) = 0.6 mag and its full width 
half maximum points to variations in E$_{\rm g}$(B-V) up to 0.4 mag across the galaxy.
The presence of a gradient in E$_{\rm g}$(B-V) along the galaxy major axis is evident in Fig. 3, 
where we plot the azimuthal E$_{\rm g}$(B-V) as a function of position along the semi-major axis 
(solid line; the dotted line represents the foreground E$_{\rm G}(B-V)$ = 0.51 mag). We integrate
the observed flux of the H$\alpha$ and Br$\gamma$ images over concentric, elliptical semi annuli and
compute the H$\alpha$/Br$\gamma$ flux ratio to estimate the azimuthal E$_{\rm g}$(B-V) 
as a function of position along the semi-major axis of NGC~1569. From now on,
we assume the convention of negative distances for increasing right ascension from the 
galaxy center. The vertical errorbar ($\pm$0.1 dex) in Fig. 3 indicates 
the systematic uncertainty on E$_{\rm g}$(B-V) due to the uncertainty 
on the flux calibration. Because of our S/N cut, the color excess can be determined only for 
galactocentric distances $<$-50 pc and $>$30 pc. As Fig. 3 shows, the color excess increases
from E$_{\rm g}$(B-V) $\simeq$ 0.57 mag in the 
South-East portion of the galaxy to E$_{\rm g}$(B-V) =  0.7 - 0.8 mag in the NW. 
This gradient is consistent with the findings of Kobulnicky \& Skillman (1997). The
higher E$_{\rm g}$(B-V) ($\sim$0.7) seen at $+$100 pc from the central cavities in Fig. 3
is in agreement with the findings of Gonz\'alez et al. (1997). 
\par
We use the 2D spatial map of E$_{\rm g}$(B-V) to correct the line-emission 
images pixel by pixel following the formula:

\begin{equation}
F_i(\lambda) = F_o(\lambda) \times 10^{0.4 {\rm E_{\rm g}(B-V)} \kappa(\lambda)}
\end{equation}

\par\noindent
where $F_i(\lambda)$ and $F_o(\lambda)$ are the intrinsic and observed line flux,
respectively, and $\kappa(\lambda)$ is the extinction law. For a Fitzpatrick's (1999)
extinction curve $\kappa$ is 2.535, 1.096, 0.562 and 0.363 for H$\alpha$, HeI 1.083$\mu$m,
[FeII] 1.64$\mu$m and Br$\gamma$, respectively (Calzetti, private communication).
The correction for dust extinction enables us to correct the [FeII] 1.64$\mu$m emission for the
contribution of the Br12 line, estimated to be 0.17 times the Br$\gamma$ flux (assuming
Case B, Storey \& Hummer 1995). We then compute the surface brightness profile of NGC~1569 for concentric,
elliptical, semi annuli of fixed ellipticity (b/a = 0.45) and fixed orientation of the semi-major axis
(P.A. = 117$^o$) in each filter. These profiles are shown in Fig. 4 where the surface
brightness is corrected for reddening but not for the galaxy inclination. The gap
in the line-emission profiles between -50 and 30 pc is due to our S/N cut
masking out the central gas-deficient regions.  
The surface-brightness profiles of the H and He lines share a similar shape, 
with maxima at a galactocentric distance of about $\pm$300 pc. At smaller galactocentric
distances the surface brightness of the H and He lines declines, while that of the [FeII]
line stays constant and is $\sim$0.6 dex brighter than Br$\gamma$. At galactocentric
distances $<$-300 pc and $>$300 pc all five profiles are close to an exponential disk with 
similar scale lengths. 
All profiles appear to be steeper in the NW half of the galaxy with a typical, exponential scale 
length of 70 pc against 90 pc as measured for the SE half. The peak at -550 pc in the [FeII] surface 
brightness is due to the emission from a compact supernova remnant.

\section{The recent star formation activity of NGC~1569}
We integrate the dereddened H$\alpha$, HeI 1.08$\mu$m and Br$\gamma$ maps of NGC~1569 over 
an ellipse of semi-major axis $a =$ 37$''$ (606 pc) and semi-minor axis $b$ = 17$''$ (274 pc)
which does not include the extended H$\alpha$ superbubbles. Pixels below our S/N cut are
excluded from the integration.
We obtain a total, dereddened luminosity of 5.31 $\times$ 10$^{40}$ erg~s$^{-1}$ ($\pm$0.20 dex),
4.51 $\times$ 10$^{39}$ erg~s$^{-1}$ ($\pm$0.14 dex) and 5.05 $\times$ 10$^{38}$ erg~s$^{-1}$ 
($\pm$0.11 dex) for the H$\alpha$, HeI 1.08$\mu$m and Br$\gamma$ lines respectively, at the adopted 
distance of 3.36 Mpc. The quoted uncertainties (all at 1$\sigma$ level) are due to flux calibration and reddening
correction. We use these values to study the stellar population responsible for ionizing the interstellar
H and He, and the dereddened H$\alpha$ and Br$\gamma$ images to trace the spatial distribution of the 
SFR density across NGC~1569.

\subsection{The recent starburst}
We compute the relation between the number of ionizing photons [$Q({\rm H^0})$] for hydrogen
and the H$\alpha$ (Br$\gamma$) line luminosity, as well as the relation between the number of
ionizing photons [$Q({\rm He^0})$] for helium and the HeI 1.08$\mu$m line luminosity, under 
Case B with a temperature of 10,000 K and a density of 100 cm$^{-3}$ (typical of HII regions):

\begin{equation}
Q({\rm H^0})~{\rm [s^{-1}]} = 7.3 \times 10^{11}~L({\rm H\alpha)~[erg~s^{-1}]}
\end{equation}

\begin{equation}
Q({\rm H^0})~{\rm [s^{-1}]} = 7.5 \times 10^{13}~L({\rm Br\gamma)~[erg~s^{-1}]}
\end{equation}

\begin{equation}
Q({\rm He^0})~{\rm [s^{-1}]} = 1.0 \times 10^{12}~L({\rm HeI~1.08\mu m)~[erg~s^{-1}]}
\end{equation}

Applying Eq.(3) - (5) to the total, dereddened luminosities from above yields
$Q({\rm H^0})$ = 3.88 $\times$ 10$^{52}$ s$^{-1}$ from the H$\alpha$ line,
$Q({\rm H^0})$ = 3.79 $\times$ 10$^{52}$ s$^{-1}$ from the Br$\gamma$ line, and  
$Q({\rm He^0})$ = 4.51 $\times$ 10$^{51}$ s$^{-1}$. In Fig. 5 we compare 
$Q({\rm H^0})$ (as derived from the Br$\gamma$ line) and $Q({\rm He^0})$
with STARBURST99 (Leitherer et al. 1999) predictions for an instantaneous
burst of star formation (panel $a$) and for continuous star formation (panel $b$). For both
scenarios we adopted $Z$ = 0.2 $Z_{\odot}$ and a Kroupa (2001) initial mass function (with exponents = 
1.3 and 2.3 for the mass ranges 0.1 - 0.5 M$_{\odot}$ and 0.5 - 100 M$_{\odot}$ respectively); 
stellar evolution was modeled with the Geneva evolutionary tracks for high mass-loss rates 
(Meynet et al. 1994). In panel $a$ of Fig. 5 the correlation between $Q({\rm H^0})$ and 
$Q({\rm He^0})$ for different stellar ages
and at fixed total mass (M$_{\rm tot}$) of the burst is traced by solid lines corresponding to 
M$_{\rm tot}$ = 10$^5$, 10$^6$ and 10$^7$ M$_{\odot}$. The dashed lines represent the correlation 
between $Q({\rm H^0})$ and $Q({\rm He^0})$ for different M$_{\rm tot}$ and at fixed stellar age 
(1, 3, 5, and 7 Myr). The number of H and He ionizing photons measured for NGC~1569 from its 
Br$\gamma$ and HeI 1.08$\mu$m lines is plotted 
with a black-filled circle whose errorbars ($\pm$1$\sigma$) take into account the systematic 
uncertainty due to flux calibration and reddening correction. This number is consistent with 
being produced by a 4 Myr ($\pm$0.3 Myr at 1$\sigma$ level) old stellar population, 
comprising $\sim$5400 O stars and $\sim$250 Wolf-Rayet stars and with a total mass of 
1.8 $\times$ 10$^6$ M$_{\odot}$ (with a 20\% uncertainty, at 1$\sigma$ level).
In panel $b$ the observed values of $Q({\rm H^0})$ and $Q({\rm He^0})$ are compared with the 
predictions for continuous star formation: the solid lines show the correlation between 
$Q({\rm H^0})$ and $Q({\rm He^0})$ at fixed SFR of 0.05, 0.10, 0.15 and 0.20 M$_{\odot}$~yr$^{-1}$, 
and their length spans the time interval between 1 Myr and 1 Gyr. At stellar ages older than 1 Gyr, 
$Q({\rm H^0})$ and $Q({\rm He^0})$ cease to change with time. None of the tracks fits within
1$\sigma$ the number of H and He ionizing photons measured for NGC~1569; the observed 
$Q({\rm H^0})$ and $Q({\rm He^0})$ are marginally consistent with
a star formation history with SFR $\geq$ 0.15 M$_{\odot}$~yr$^{-1}$ that has been going on for
more than 10 Myr.

We checked whether the properties of the stellar population ionizing the H and He gas
in NGC~1569 vary with galactocentric distance. We integrated the flux of the dereddened H$\alpha$ and 
HeI 1.083$\mu$m images over concentric, elliptical semi annuli and derived the corresponding 
$Q({\rm H^0})$ and $Q({\rm He^0})$
as a function of galactocentric distance. We compared these $Q({\rm H^0})$ and $Q({\rm He^0})$ values
with the SB99 predictions for instantaneous bursts, and found that stellar ages are consistent with a 
value of $\sim$4 Myr at any galactocentric distance. On the other hand, the mass surface density of the ionizing stars increases inward, from $\sim$0.01 M$_{\odot}$~pc$^{-2}$ at $\pm$700 pc from the galaxy center to 
$\sim$0.3 M$_{\odot}$~pc$^{-2}$ at 300 pc in the SE portion of NGC~1569 and to $\sim$1.8 M$_{\odot}$~pc$^{-2}$ 
in the North-West.

\subsection{The star formation rate and its density}
Using the relation between star formation rate (SFR) and the dereddened
H$\alpha$ (or Br$\gamma$) luminosity as in Kennicutt (1998) we derive an average SFR of 
$\simeq$0.4 M$_{\odot}$~yr$^{-1}$, consistent with what can be inferred from the age and total
mass of the burst derived above. The uncertainty on the flux calibration of both 
emission lines and on the reddening determination gives rise to a systematic uncertainty of 
$\pm$0.20 dex on the SFR derived from the H$\alpha$ (or 0.11 dex on the SFR estimated from
the Br$\gamma$).
\par
The 2D map of the star formation rate per pixel, obtained from the dereddened Br$\gamma$ image, 
is shown in Fig. 6 in logarithmic units; the SFR across NGC~1569 is typically 
$\simeq$10$^{-6}$ M$_{\odot}$~yr$^{-1}$ per pixel (i.e. 3 $\times$ 10$^{-7}$ M$_{\odot}$~yr$^{-1}$~pc$^{-2}$)
and increases to few 10$^{-5}$ M$_{\odot}$~yr$^{-1}$ per pixel (i.e. $\simeq$ 10$^{-5}$ 
M$_{\odot}$~yr$^{-1}$~pc$^{-2}$) at each side of the cavities where the H and He line
emissions peak. This trend is well described by the azimuthal SFR density (obtained by integrating the 
dereddened flux of the Br$\gamma$ image over concentric, elliptical semi annuli) plotted as a function of 
position along the semi-major axis (see Fig. 7). The maxima of the SFR density are about 9 $\times$ 10$^{-7}$
M$_{\odot}$~yr$^{-1}$~pc$^{-2}$ and 4 $\times$ 10$^{-6}$ M$_{\odot}$~yr$^{-1}$~pc$^{-2}$ at -300 pc and
+300 pc from the galaxy center, respectively. The errorbars ($\pm$1$\sigma$) in Fig. 7 indicates a 
systematic uncertainty of $\pm$0.1 dex on the SFR derived from the dereddened Br$\gamma$ image.
\par
The width of the distribution of the pixel star formation rate as obtained from the dereddened Br$\gamma$
image indicates that SFR density varies by a factor of 10 within the galaxy.

\section{The [FeII] emission}
The [FeII] emission, with its prominent lines at 1.26 $\mu$m and 1.64 $\mu$m, is usually explained as
triggered by electron collisions occurring in a zone of partially ionized hydrogen where Fe$^{+}$ and
$e^{-}$ coexist. The emission intensity is proportional to the size of this region, which
is rather small in HII regions but extended in supernova remnants (SNRs, cf. Oliva, Moorwood \&
Danziger 1989). The [FeII] emission would thus be a clear signpost for SNRs, and as such 
it is expected to come from compact sources in external galaxies. Recent studies have shown that
the [FeII] emission can also be spatially extended in galaxies known to experience a galactic 
wind (e.g. NGC~253, Forbes et al. 1993, Sugai, Davies \& Ward 2003; NGC~5253, Cresci et al. 2010, 
Labrie \& Pritchet 2006, Turner, Beck \& Ho 2000; M~82, Greenhouse et al. 1997, Heckman et al. 1987).
Some theoretical computations (e.g. Seab \& Shull 1983 and McKee et al. 1987) have shown that
this extended [FeII] emission can be explained by high-speed shocks such as those associated with
a galactic outflow. The proposed mechanism assumes that iron atoms can be extracted from silicate
grains that are preferentially destroyed via nonthermal sputtering and grain-grain
collisions (cf. also McCarthy, Heckman \& van Breugel 1987; van der Werf et al. 1993. We note that these 
mechanisms can also be triggered by SN shocks). 
NGC~1569 is yet another starburst galaxy exhibiting spatially extended 
emission of [FeII] 1.64$\mu$m (already detected by Labrie \& Pritchet) and a galactic outflow 
(Heckman et al. 1995).
\par
Given that the Br$\gamma$ emission comes from gas ionized by OB stars, the flux ratio 
[FeII] 1.64$\mu$m/Br$\gamma$ 
is commonly used to trace the number ratio of supernovae (SNs) to OB stars, hence the star formation 
history of a galaxy at fixed initial mass function. Values of this ratio larger than few tens indicate
that excitation is mostly produced by SN shocks, while values $<<$ 1, typically 0.1 or lower, are
due to photoionization (Alonso-Herrero et al. 1997). The dereddened [FeII] 1.64$\mu$m luminosity (already 
corrected for the emission of the Br12 line) integrated over an ellipse of semi-major axis $a =$ 37$''$ 
and semi-minor axis $b$ = 17$''$ (excluding the central cavities) is 4.67 $\times$ 10$^{38}$ erg s$^{-1}$ at the adopted distance of 3.36 Mpc.
From this and the integrated luminosity of Br$\gamma$, we obtain a global [FeII] 1.64$\mu$m/Br$\gamma$ ratio 
= 0.9, suggesting that excitation by SN shocks may prevail over photoionization by OB stars formed 
during the most recent burst. Figure 8 shows the two-dimensional spatial distribution of the 
[FeII] 1.64$\mu$m/Br$\gamma$ ratio across NGC~1569. The line ratio is larger than 1 
mainly in the galaxy outskirts and around the central cavities, and it is $<$ 1 especially 
in two regions at the opposite extremes of the cavities coincident with the peaks
of the H$\alpha$ emission and the star formation rate density. Another way of looking at the spatial 
variation of the [FeII] 1.64$\mu$m/Br$\gamma$ ratio is to integrate the dereddened flux of the [FeII] 1.64$\mu$m 
and Br$\gamma$ images over concentric, elliptical semi annuli and compute the [FeII] 1.64$\mu$m/Br$\gamma$
flux ratio as a function of position along the semi-major axis. This is plotted in Fig. 9; the minima at [FeII] 
1.64$\mu$m/Br$\gamma <$ 1 occur
at about $\pm$300 pc from the galaxy center where the star formation rate density is highest (cf. Fig.
8). The peak at -550 pc is due to a compact SNR. The [FeII] 1.64$\mu$m/Br$\gamma$ ratio increases above 1 with galactocentric distance and at the edges of the cavities, possibly
suggesting that in these regions star formation has been somehow quenched. A relevant caveat on this last 
interpretation comes from the possibility that iron has been excited by shocks produced by the galactic
outflow of NGC~1569. In this case, the [FeII] 1.64$\mu$m/Br$\gamma$ ratio may be expected not to trace the 
star formation history of a galaxy on small spatial scales. The extent by which a galactic outflow can alter 
this ratio so that it is no longer representative of the local number ratio SNs vs OB stars is still to be 
thoroughly investigated. In their spectroscopic study of the central 20$'' \times$ 20$''$ in
NGC~1569 Westmoquette et al. (2007) found little evidence for shocked line ratios, most likely because
this area is experiencing intense star formation and photoionization from OB stars. In this region
(equivalent to about 320 $\times$ 320 pc$^2$ at the adopted distance of 3.36 Mpc) we measure the lowest
values of [FeII] 1.64$\mu$m/Br$\gamma$ (see Fig. 9, exception made for the cavities edge), which may also 
be interpreted as due to photoionization.
\par
We used the dereddened, integrated luminosity of the [FeII] 1.64$\mu$m and Br$\gamma$ lines to
estimate the supernova rate (SNR) and the gas-phase abundance of Fe$^+$ for NGC~1569 as a whole.
Under the assumption that the typical [FeII] 1.64$\mu$m luminosity of a supernova is $\sim$10$^{37}$ 
erg~s$^{-1}$ and its average life $\sim$10$^4$ yr (cf. Lumsden \& Puxley 1995), we estimate a SNR 
of $\sim$0.005 yr$^{-1}$ for the whole galaxy. As for the gas-phase abundance of Fe$^+$, we adopt
Greenhouse et al.'s (1997) prescription, where:

\begin{equation}
\frac{N({\rm Fe^+})}{N({\rm H})} \sim 9 \times 10^{-6} \frac{L({\rm [FeII] 1.64})}{L({\rm Pa} \beta)}
\end{equation}
\par\noindent
under Case B for a temperature of 10,000 K and a density of 100 cm$^{-3}$. For the same physical
conditions, $L$(Pa$\beta$) = 5.86~$L$(Br$\gamma$), hence:

\begin{equation}
\frac{N({\rm Fe^+})}{N({\rm H})} \sim 1.54 \times 10^{-6} \frac{L({\rm [FeII] 1.64})}{L({\rm Br} \gamma)}
\end{equation}
\par\noindent

We thus derive $N{\rm (Fe^+)}/N{\rm (H)} \geq$ 1.4 $\times$ 10$^{-6}$ and $N{\rm (Fe)}/N{\rm (Fe)_{\odot}} 
\geq$ 0.05, where the Sun $N{\rm (Fe)}/N{\rm (H)}$ is 2.8 $\times$ 10$^{-5}$ (Holweger 2001).  
The lower limit comes from the assumption that all the iron is singly ionized.

\section{Conclusions}
The wide field of view and high angular resolution of the infrared imager/spectrograph LUCIFER
mounted on the Large Binocular Telescope allowed us to acquire deep and detailed images of the
local starburst NGC~1569 in the HeI 1.08$\mu$m, [FeII] 1.64$\mu$m and Br$\gamma$ light. Together
with $HST$/ACS H$\alpha$ images of the galaxy, these data were used to derive the two-dimensional
spatial distributions (on scales as small as 2 pc) of dust extinction (foreground $+$ intrinsic),
star formation rate density and [FeII] 1.64$\mu$m/Br$\gamma$ ratio, and to estimate the age and total mass
of the most recent burst responsible for ionizing the interstellar medium (ISM) in NGC~1569. The galaxy looks
clearly asymmetric along its major axis with respect to its center, where its super-star clusters
A and B have most likely evacuated gas and formed gas-deficient cavities. The color excess E$_{\rm g}$(B-V), 
as derived from the H$\alpha$/Br$\gamma$ ratio, is patchy 
with a scatter of about 0.4 mag across the galaxy. Reddening is lower in the SE portion of NGC~1569 
with an average E$_{\rm g}$(B-V) $\simeq$ 0.57 mag and increases to 0.7 - 0.8 mag in the NW along the galaxy major 
axis. Similarly to the surface brightness profiles of the H and He line emissions, the star formation 
rate density increases inward along the major axis, reaching two maxima of 9 $\times$ 10$^{-7}$ and 
4 $\times$ 10$^{-6}$ M$_{\odot}$~yr$^{-1}$~pc$^{-2}$ at -300 and +300 pc, respectively, from the 
galaxy center (i.e. just outside the central cavities). Within the galaxy the star formation rate 
density is seen to vary by a factor of 10. 
The peaks in SFR density are spatially coincident with the minima in the [FeII] 1.64$\mu$m/Br$\gamma$ 
flux ratio, suggesting that the OB stars formed during the most recent burst are responsible for
the ISM photoionization and are preferentially located in a ring around the central cavities.
The [FeII] 1.64$\mu$m/Br$\gamma$ ratio is seen increasing outward from the galaxy center and along its
major axis, as well as at the edges of the central cavities. At face value, this trend may imply 
that excitation by SN shocks dominates over photoionization from OB stars and indicate that star
formation has been quenched some Myrs ago in the central cavities and in the outermost regions 
of the galaxy.
This trend would be consistent with the findings of Aloisi et al. (2001), whereby the age of stars
increases with galactocentric distance from 50 Myr close to the galaxy center to $>$1 Gyr in the
galaxy outskirts.
How and to which extent the intervention of a galactic outflow as in NGC~1569 alters the 
[FeII] 1.64$\mu$m/Br$\gamma$ 
ratio and prevents it from tracing the ISM excitation mechanisms on small spatial scales remains to
be fully investigated. On global scales, the [FeII] 1.64$\mu$m luminosity integrated over the galaxy main body
points to a SN rate of about 0.005 yr$^{-1}$ and an iron abundance $N{\rm (Fe^+)}/N{\rm (H)} \geq$ 1.4 
$\times$ 10$^{-6}$. This value is quite consistent with the Fe abundance in the ISM of the Small Magellanic
Cloud (Rolleston et al. 2003).
If the high-speed shocks in the galactic outflow would be responsible for extracting iron atoms from silicate
grains (as suggested by the theoretical calculations of, for example, Seab \& Shull 1983 and McKee et al. 1987)
and hence for a spatially extended [FeII] emission, one would expect to find the ISM Fe abundance in NGC~1569 
to be enhanced with respect to the Small Magellanic Cloud (SMC) which has a stellar metallicity similar to 
NGC~1569 but no galactic outflow. Given that the ISM Fe abundance is similar in both NGC~1569 and the SMC, the 
present data would tend to exclude that the diffuse [FeII] emission in NGC~1569 is due to the galactic outflow. 
Moreover, Sofia et al. (2006) found that iron atoms in the ISM of the SMC are most likely locked in metal 
grains and/or oxides rather than in silicates. If this applied to NGC~1569 also, the galactic outflow would not be 
able to extract iron from dust grains (as in the scenario proposed by Seab \& Shull 1983 and McKee et al. 1987), increase the Fe emission and thus explain its spatial extension.
\par
We used the H$\alpha$, Br$\gamma$ and HeI 1.08$\mu$m luminosities integrated over NGC~1569 to estimate
the star formation rate of the most recent burst ($\simeq$0.4 M$_{\odot}$~yr$^{-1}$) and the number
of stellar photons able to ionize the interstellar H and He gas [i.e. $Q({\rm H^0})$ and $Q({\rm He^0})$].
When compared to STARBURST99 predictions the observed $Q({\rm H^0})$ and $Q({\rm He^0})$ enable us to
constrain the age and the mass of the stellar population responsible for ionizing H and He.
In the case of NGC~1569 the values of $Q({\rm H^0})$ and $Q({\rm He^0})$ are in good agreement with
an instantaneous burst of star formation. A scenario where star formation is continuous and takes place 
with a constant SFR is not rejected by statistics, but is seemingly inconsistent with the gasping nature 
of the star formation activity undergone by NGC~1569 in the last 1 Gyr, 
as derived by Angeretti et al. (2005) from the color-magnitude diagram of resolved stars. The best-fitting 
instantaneous burst occurred about 4 Myr ago and involved a total stellar mass of 
$\simeq$1.8 x 10$^6$ M$_{\odot}$. It gave birth to about 5400 O stars and 250 Wolf-Rayet stars. 
Such a stellar age is constant with galactocentric distance and where H$\alpha$, Br$\gamma$ and HeI 1.08$\mu$m
emissions are detected. The only property of the burst changing with distance from the galaxy center
is the surface mass density of the ionizing stars, which follows the surface brightness profiles of 
the H and He lines and increases from $\sim$0.01 M$_{\odot}$~pc$^2$ in the galaxy outskirts up to 
$\sim$1.8 M$_{\odot}$~pc$^2$ just outside the central cavities.
The comparison of this latest starburst with the three episodes of star formation identified by
Angeretti et al. (2005, cf. Sect. 1) suggests a star formation history where the burst duration and average
SFR have become shorter and larger, respectively, from 1 Gyr ago to 4 Myr ago. Consequently, the total mass 
in stars produced by these bursts of star formation has decreased by a factor of about 20 in the last 1
Gyr.
Interestingly enough, the 4 Myr age derived from the observed $Q({\rm H^0})$ and $Q({\rm He^0})$ is close 
to that of the younger stellar component found mostly in SSC A by Gonz\'alez Delgado et al. (1997). 
Therefore, the big picture emerging from these results is one where the strong stellar winds and
supernova explosions from the older stellar population of SSCs A and B (about 9 Myr old) removed a large
fraction of gas from the clusters surroundings and triggered, $\sim$4 Myr later, star formation at the 
edges of the central cavities. The gas left over in the vicinity of SSC A (and perhaps also B) underwent 
star formation nearly at the same time.

\acknowledgments
We would like to thank Daniela Calzetti and Monica Tosi for very useful discussions. 
The Dark Cosmology Centre is funded by the DNRF.

\par\noindent
{\it Facilities:}  \facility{LBT (LUCIFER)}, \facility{HST (ACS)}

\clearpage
\begin{figure}
\includegraphics[angle=90,scale=0.8]{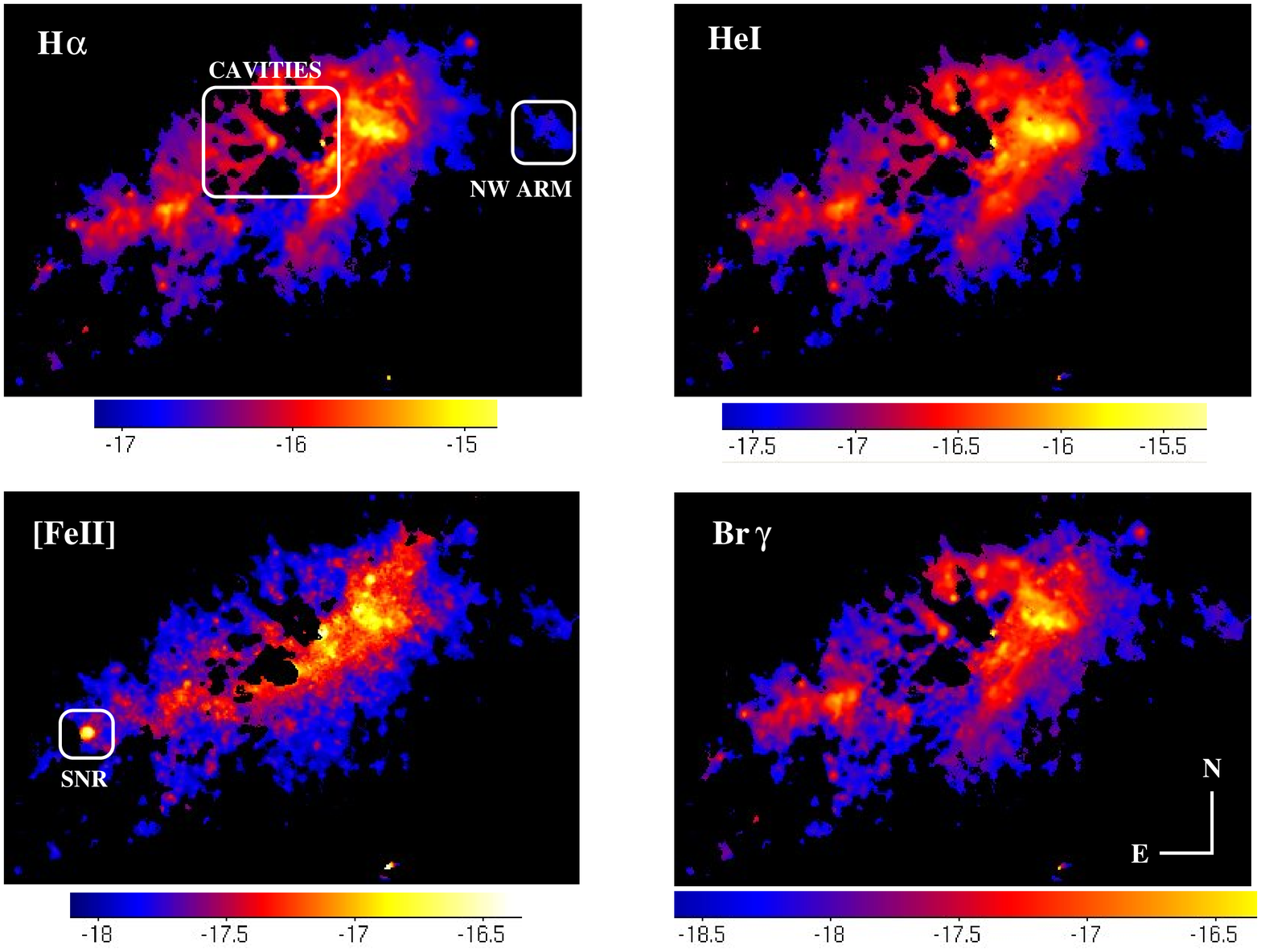}
\caption{The line-emission images of NGC~1569 calibrated in flux in units of erg~cm$^{-2}$~s$^{-1}$, but not
corrected for reddening. Fluxes are on logarithmic scale, and the image
size is 1.5 kpc $\times$ 1 kpc at the adopted distance of 3.36 Mpc. The white frames indicate the
location of the north-western (NW) arm, the central cavities and the supernova remnant (SNR).}
\end{figure}

\clearpage
\begin{figure}
\epsscale{0.9}
\plotone{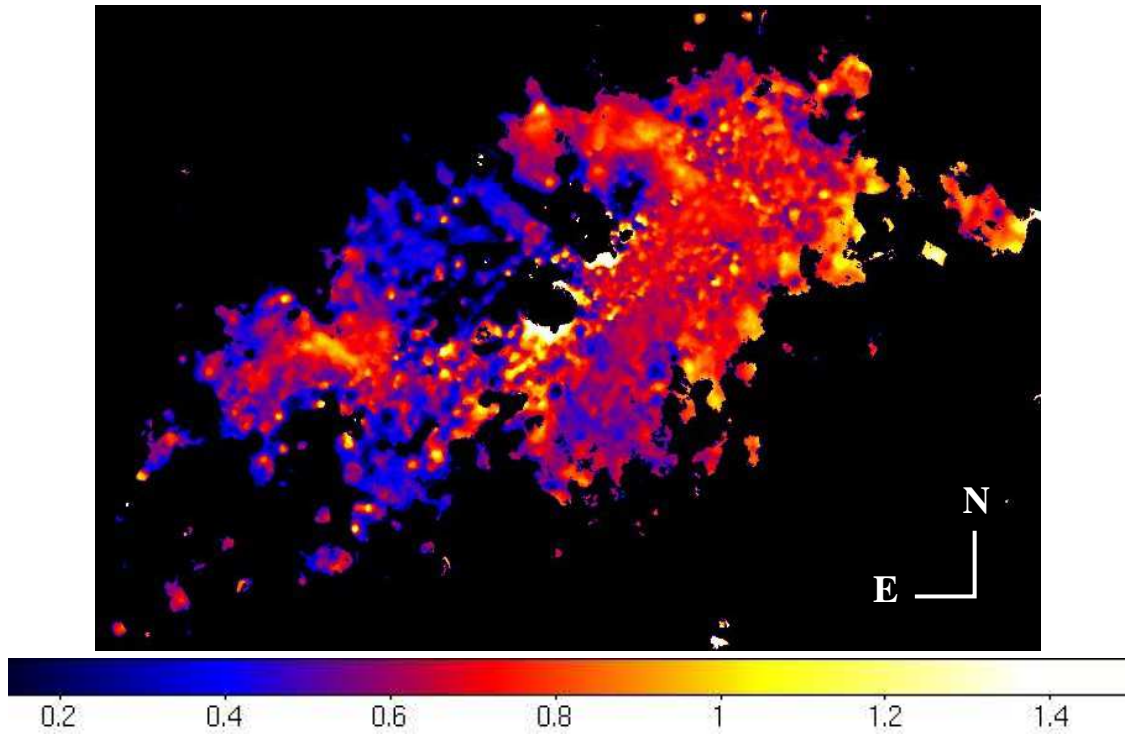}
\caption{The E$_{\rm g}$(B-V) (intrinsic $+$ foreground) map of NGC~1569 derived from
the H$\alpha$/Br$\gamma$ flux ratio. The image size
is 1.5 kpc $\times$ 1 kpc.}
\end{figure}

\clearpage
\begin{figure}
\epsscale{0.6}
\plotone{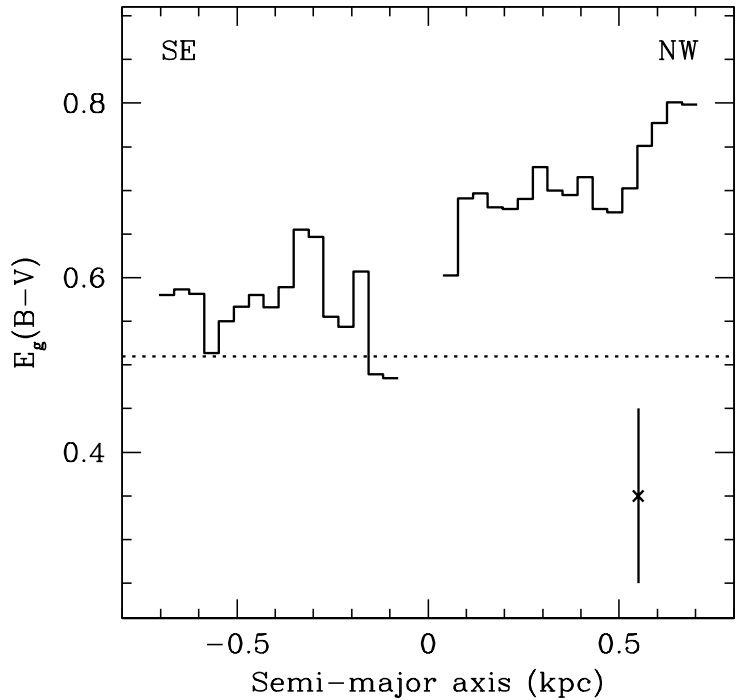}
\caption{The azimuthal E$_{\rm g}$(B-V)
(intrinsic $+$ foreground), computed over 180 degrees, as a function of
position along the semi-major axis (solid line). The dotted line
represents the foreground Galactic color excess in the direction
of NGC~1569, E$_{\rm G}$(B-V) = 0.51 mag. The errorbar corresponds to a systematic 
uncertainty on E$_{\rm g}$(B-V) of $\pm$0.1 mag, and is due to the uncertainty on the 
flux calibration. The gap between -50 and 30 pc is due to our S/N cut which masked out the central, 
gas-deficient regions.}
\end{figure}

\clearpage
\begin{figure}
\epsscale{0.6}
\plotone{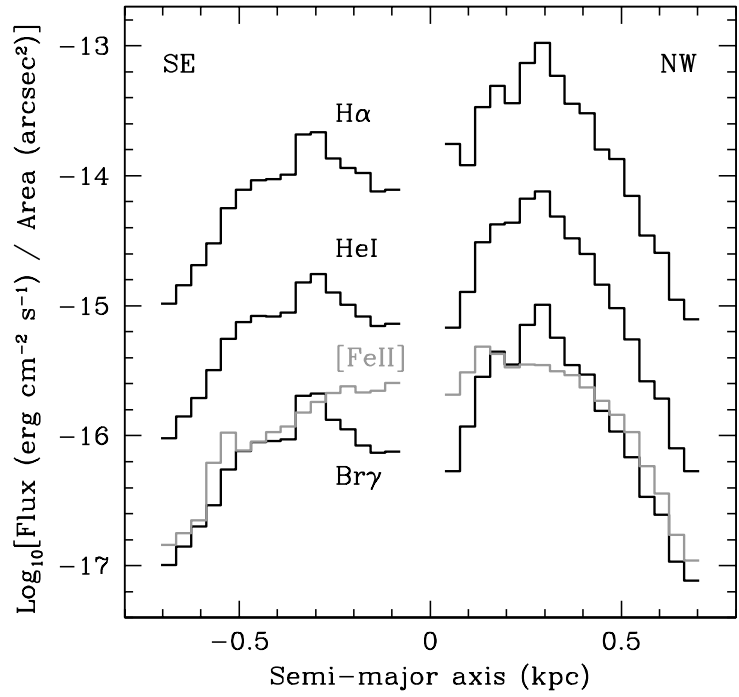}
\caption{The surface brightness profile of NGC~1569 in the different 
narrow-band filters. Fluxes are corrected for E$_{\rm g}$(B-V).
The gap between -50 and 30 pc is due to our S/N cut which 
masked out the central, gas-deficient regions.}
\end{figure}

\clearpage
\begin{figure}
\epsscale{0.9}
\plotone{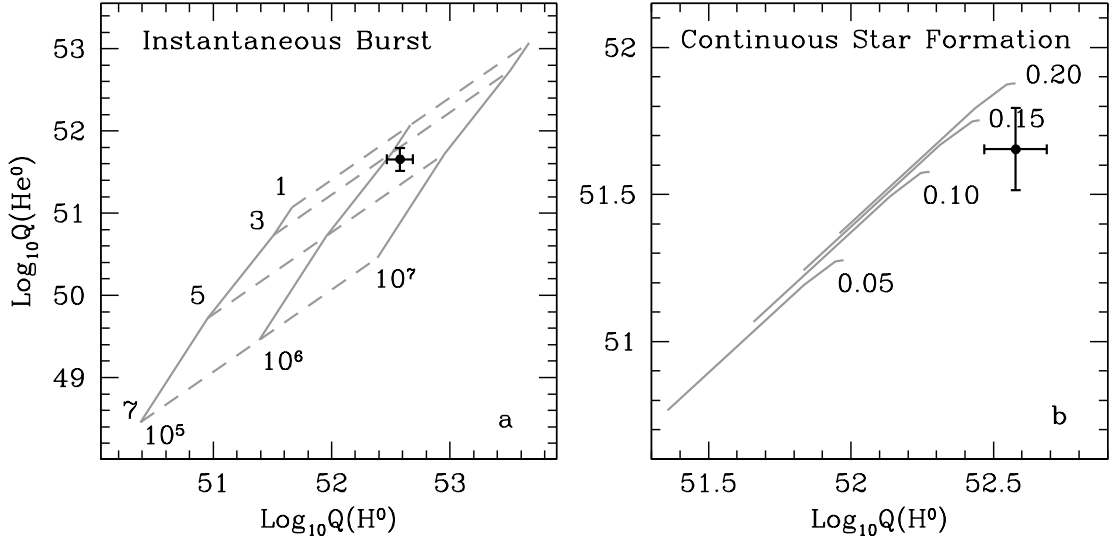}
\caption{{\it Panel a:} the number of photons ionizing He [$Q({\rm He^0})$] is plotted 
as a function of the number of photons ionizing H [$Q({\rm H^0})$] as derived from
the Br$\gamma$ line. The solid lines trace 
the correlation between $Q({\rm H^0})$ and $Q({\rm He^0})$ at fixed total mass of the burst 
(10$^5$, 10$^6$ and 10$^7$ M$_{\odot}$) and as a function of age (1, 3, 5, 7 and 
10 Myr). The dashed lines show the correlation between $Q({\rm H^0})$ and 
$Q({\rm He^0})$ at fixed age and as a function of total mass of the burst. The black-filled circle 
represents the $Q({\rm H^0})$ and $Q({\rm He^0})$ values measured for NGC~1569 in this work.
The errorbars are $\pm$1$\sigma$. 
{\it Panel b:} as before, but the solid lines here trace the predictions for 
a continuous star formation with different star formation rates (0.05, 0.10, 0.15
and 0.20 M$_{\odot}$~yr$^{-1}$). The line length corresponds to a time interval
between 1 Myr (bottom left corner of each line) and 1 Gyr (top right corner);
for ages older than 1 Gyr $Q({\rm H^0})$ and $Q({\rm He^0})$ remain unchanged.}
\end{figure}

\clearpage

\begin{figure}
\epsscale{0.7}
\plotone{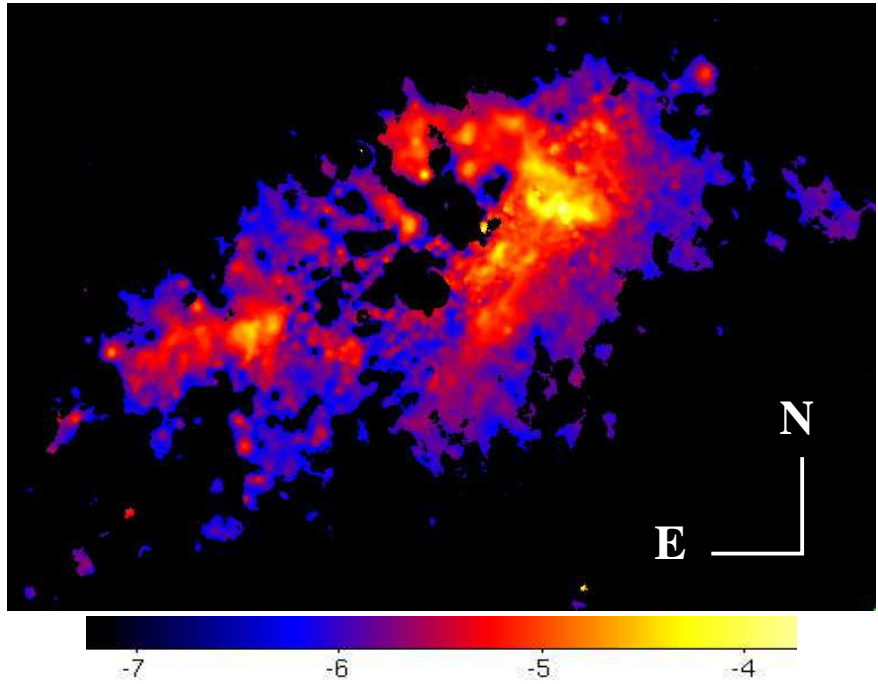}
\caption{The star formation rate map of NGC~1569 derived from the dereddened Br$\gamma$ image.
The image size is 1.5 kpc $\times$ 1 kpc. The star formation rate is in units of
M$_{\odot}$~yr$^{-1}$ per pixel, and is displayed on logarithmic scale.}
\end{figure}

\clearpage

\begin{figure}
\epsscale{0.6}
\plotone{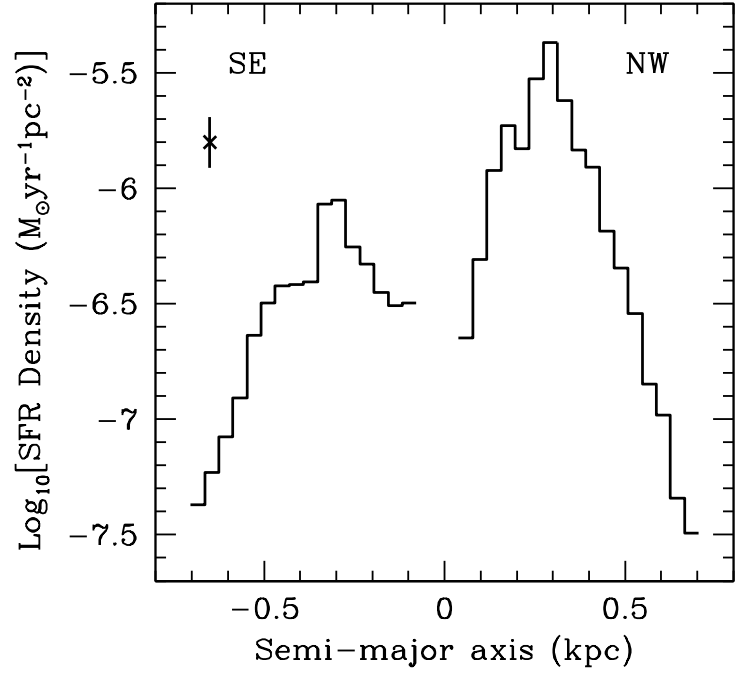}
\caption{The azimuthal SFR density as a function of position along the semi-major axis, 
obtained from the dereddened Br$\gamma$ image. 
The errorbar ($\pm$ 0.11 dex) represents the systematic uncertainty on SFR 
due to the uncertainty on the flux calibration and color excess. The gap between -50 and 30 
pc is due to our S/N cut which masked out the central, gas-deficient regions.}
\end{figure}

\clearpage

\begin{figure}
\epsscale{0.7}
\plotone{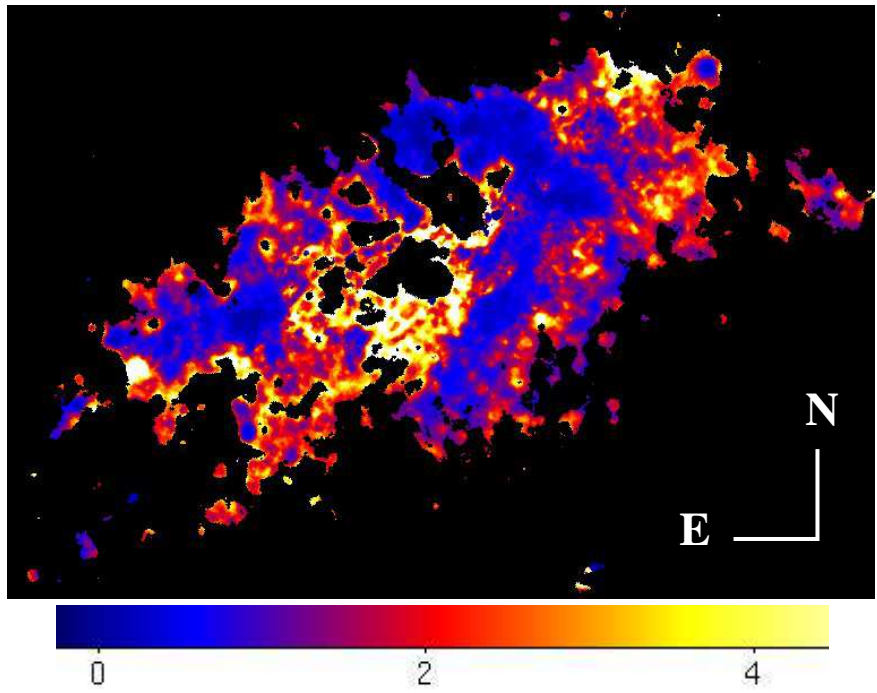}
\caption{The 2D map of the [FeII] 1.64$\mu$m/Br$\gamma$ flux ratio, after correction
for reddening. The image size is 1.5 kpc $\times$ 1 kpc.}
\end{figure}

\clearpage

\begin{figure}
\epsscale{0.6}
\plotone{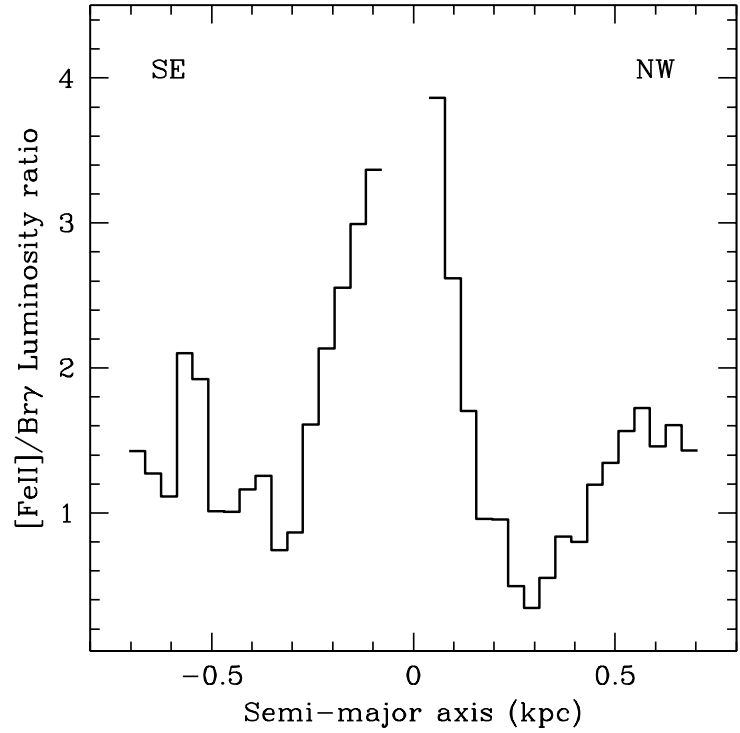}
\caption{The azimuthal [FeII] 1.64$\mu$m/Br$\gamma$ flux ratio, corrected
for reddening, as a function 
of position along the semi-major axis. The gap between -50 and 30 
pc is due to our smoothing procedure which masked out the central, gas-deficient regions.}
\end{figure}

\clearpage
\begin{table}
\begin{center}
\caption{Near-IR and optical total exposure times on source}
\begin{tabular}{cc}
\tableline\tableline
Filter   & Exposure Time (min)\\
\tableline
J                &  20 \\
H                &  55 \\
K                & 121 \\
HeI 1.08$\mu$m    &  30 \\
$\rm [FeII]$ 1.64$\mu$m &  67 \\
Br$\gamma$       & 111 \\
                 & \\
F606W            & 326 \\
F658N            &  77 \\
\tableline
\end{tabular}
\end{center}
\end{table}


\begin{thebibliography}{}
\bibitem[]{} Ageorges, N., et al., 2010, Proc. SPIE, 7735, 77351L
\bibitem[]{} Aloisi, A., et al., 2001, \apj, 121, 1425
\bibitem[]{} Alonso-Herrero, A., Rieke, M.J., Rieke, G.H., \& Ruiz, M., 1997, \apj, 482, 747
\bibitem[]{} Angeretti, L., Tosi, M., Greggio, L., Sabbi, E., Aloisi, A., 
\& Leitherer, C., 2005, \aj, 129, 2203
\bibitem[]{} Arp, H., \& Sandage, A., 1985, MPA Rep., 173, 29
\bibitem[]{} Broadhurst, T., Ellis, R., \& Glazebrook, K. 1992, Nature, 355, 55
\bibitem[]{} Burstein, D., \& Heiles, C., 1982, AJ, 87, 1165
\bibitem[]{} Calzetti, D.,  Kinney, A., \& Storchi-Bergmann, T., 1994, \apj, 429, 582
\bibitem[]{} Calzetti, D.,  Kinney, A., \& Storchi-Bergmann, T., 1996, \apj, 458, 132
\bibitem[]{} Cole, S., Arag\'on-Salamanca, A., Frenk, C.S., Navarro, J.F., \&
Zepft, S.E. 1994, \mnras, 271, 781
\bibitem[]{} Cresci, G., Vanzi, L., Sauvage, M., Santangelo, G., \& van der Werf, P., 2010,
submitted to A\&A, arXiv:1007.3666
\bibitem[]{} De Marchi, G., Clampin, M., Greggio, L., Leitherer, C., Nota, A., \& Tosi, M.,
1997, \apj, 479, L27
\bibitem[]{} Fitzpatrick, E.L., 1999, PASP, 111, 63
\bibitem[]{} Forbes, D., Ward, M.J., Rotaciuc, V., Blietz, M., Genzel, R., Drapatz, S., van der Werf,
P.P., \& Krabbe, A., 1993, \apj, 406, L11
\bibitem[]{} Gonz\'alez Delgado, R.M., Leitherer, C., Heckman, T., \& Cervi\~no, M., 1997, \apj, 483, 705
\bibitem[]{} Greenhouse, M.A., et al., 1997, \apj, 476, 105
\bibitem[]{} Greggio, L., Tosi, M., Clampin, M., De Marchi, G., Leitherer, C., Nota, A., \&
Sirianni, M., 1998, \apj, 504, 725
\bibitem[]{} Grocholski, A.J., et al., 2008, \apj, 686, L79
\bibitem[]{} Heckman, T.M., Armus, L., McCarthy, P., van Breugel, W., \& Miley, G.K., 1987, in Star
formation in galaxies. ed. C.J. Lonsdale Persson, p. 461 (N87-24266 17-89)
\bibitem[]{} Heckman, T.M., Dahlem, M., Lehnert, M.D., Fabbiano, G., Gilmore, D., \& Waller, W.H., 1995,
\apj, 448, 98
\bibitem[]{} Hill, J.M., Green, R.F., \& Slagle, J.H., 2006, Proc. SPIE, 6267, 62670Y
\bibitem[]{} Ho, L.C., \& Filippenko, A.V., 1996, \apj, 466, L83
\bibitem[]{} Holweger, H., 2001, in Solar and Galactic Composition: A Joint SOHO/ACE Workshop, AIP
Conference Proceedings, 598, 23
\bibitem[]{} Hunter, D.A., Hawley, W.N., Gallagher, J.S., III, 1993, \aj, 106,1797
\bibitem[]{} Izotov, Y.I., \& Thuan, T.X. 1999, \apj, 511, 639
\bibitem[]{} Kauffmann, G., White, S.D.M., \& Guiderdoni, B. 1993, \mnras,
264, 201 
\bibitem[]{} Kennicutt, R.C., Jr., 1998, ARA\&A, 36, 189
\bibitem[]{} Kobulnicky, H.A., \& Skillman, E.D., 1997, \apj, 489, 636 
\bibitem[]{} Kroupa, P., 2001, \mnras, 322, 231
\bibitem[]{} Labrie, K., \& Pritchet, C.J., 2006, \apjs, 166, 188
\bibitem[]{} Leitherer, C., et al., 1999, \apjs, 123, 3
\bibitem[]{} Lilly, S.S.J., Tresse, L., Hammer, F., Crampton, D., \& 
Le Fevre, O., 1995, \apj, 455, 108
\bibitem[]{} Lumsden, S.L., \& Puxley, P.J., 1995, \mnras, 276, 723
\bibitem[]{} Martin, C.L., 1998, \apj, 506, 222
\bibitem[]{} McCarthy, P.J., Heckman, T.M., \& van Breugel, W., 1987, \aj, 93, 264
\bibitem[]{} McKee, C.F., Hollenbach, D.J., Seab, C.G., \& Tielens, A.G.G.M., 1987, \apj, 318, 674
\bibitem[]{} Meurer, G., 1995, Nature, 375, 742
\bibitem[]{} Meurer, G., Heckman, T.M., Leitherer, C., Kinney, A., Robert, C., \&
Garnett, D., 1995, \aj, 110, 2665
\bibitem[]{} Meynet, G., Maeder, A., Schaller, D., \& Charbonnel, C., 1994, A\&AS, 103, 97
\bibitem[]{} Moustakas, J., \& Kennicutt, R.C., Jr, 2006, \apjs, 164, 81
\bibitem[]{} Oliva, E., Moorwood, A.F., \& Danziger, I.J., 1989, A\&A, 214, 307
\bibitem[]{} Origlia, L., Leitherer, C., Aloisi, A., Greggio, L., \& Tosi, M., 2001, 
\aj, 122, 815
\bibitem[]{} Osterbrock, D.E., 1989, Astrophysics of Gaseous Nebulae and Active Galactic Nuclei,
University Science Books: Mill Valley, California
\bibitem[]{} Reakes, M., 1980, \mnras, 192, 297
\bibitem[]{} Rolleston, W.R.J., Venn, K., Tolstoy, E., \& Dufton, P.L., 2003, A\&A, 400, 21
\bibitem[]{} Seab, C.G., \& Shull, J.M., 1983, 275, 652
\bibitem[]{} Seifert, W., et al., 2010, Proc. SPIE, 7735, 77357W
\bibitem[]{} Stil, J.M., \& Israel, F.P., 1998, A\&A, 337, 64
\bibitem[]{} Storey, P.J., \& Hummer, D.G., 1995, MNRAS, 272, 41
\bibitem[]{} Sugai, H., Davies, R.I., \& Ward, M.J., 2003, \apj, 584, 9
\bibitem[]{} Turner, J.L., Beck, S.C., \& Ho, P.T.P., 2000, \apj, 532, 109 
\bibitem[]{} Vallenari, A., \& Bomans, D.J., 1996, A\&A, 313, 713
\bibitem[]{} van den Bergh, S., 1995, Nature, 374, 215
\bibitem[]{} van der Werf, P.P., Genzel, R., Krabbe, A., Blietz, M., Lutz, D., Drapatz, S., Ward, M.J.,
\& Forbes, D.A., 1993, \apj, 405, 522
\bibitem[]{} Waller, W.H., 1991, \apj, 370, 144
\bibitem[]{} Westmoquette, M.S., Smith, L.J., \& Gallagher, J.S., III, 2008, \mnras, 383, 864 
\bibitem[]{} Westmoquette, M.S., Smith, L.J., Gallagher, J.S., III, \& Exter, K.M., 2007, \mnras, 381, 913
\bibitem[]{} White, S.D.M., \& Frenk, C.S. 1991, \apj, 379, 52
\bibitem[]{} Whitmore, B.C., et al., 1993, \aj, 106, 1354 
\bibitem[]{} Zibetti, S., 2009, submitted to MNRAS (arXiv0911.4956)
\bibitem[]{} Zibetti, S., Charlot, S., \& Rix, H.-W., 2009, MNRAS, 400, 1181

\end{thebibliography}
\end{document}